\documentclass{elsart}
\usepackage{graphicx}
\begin{document}
\begin{frontmatter}
\title{Noise influence on solid-liquid transition of ultrathin lubricant film}
\author{Alexei V. Khomenko\thanksref{AH}}
\address{Physical Electronics Department, Sumy State
University, Rimskii-Korsakov St. 2, 40007 Sumy, Ukraine}
\thanks[AH]{E-mail: khom@phe.sumdu.edu.ua}
\date{\today}
\begin{abstract}
The melting of ultrathin lubricant film by friction between
atomically flat surfaces is studied. The additive noises of the elastic
shear stress and strain,
and the temperature are introduced for building a
phase diagram with the domains of sliding, stick-slip,
and dry friction. It is shown that increase of
the strain noise intensity causes the lubricant film melting even at
low temperatures of the friction surfaces.

\end{abstract}
\end{frontmatter}

\section{Introduction}\label{sec:level1}

The study of the noise influence on the friction process has an evident
fundamental and practical importance because in some experimental
situations the fluctuations can change the frictional behavior critically,
for example,
providing the conditions for low friction \cite{1lit} --- \cite{inlit}.
In particular, the thermal noise, acting in any experiments, can convert
the ultrathin lubricant film from stable solidlike phase state to the liquidlike one
and, thus, transform the dry friction into the sliding or the stick-slip
(the interrupted) modes.
Therefore, in recent years the considerable study has been given to the
influence of disorder and random impurities in the interface
on the static and the dynamic frictional phenomena \cite{7lit} --- \cite{8lit}.
These investigations show that a periodic surfaces are characterized
by smaller friction coefficient during sliding than nonregular ones.
Besides, the stick-slip dynamics,
inherent in solid friction, attracts an increased attention on the atomic
\cite{Yosh} --- \cite{3lit} and the macroscopic \cite{5lit,Aranson}
levels as well as for granular mediums \cite{4lit} --- \cite{aran1}.
In order to achieve the better understanding of the above phenomena, here
an analytic approach is put forward, which describes the transitions
between friction modes due to
variation of fluctuations of elastic and thermal fields.

In the previous work \cite{liq} on the basis of rheological description of
viscoelastic medium the system of kinetic
equations has been obtained, which define the mutually coordinated evolution
of the elastic shear components of the stress $\sigma$ and the strain
$\varepsilon$, and the temperature $T$ in ultrathin lubricant film during
friction between atomically flat mica surfaces. Let us write these
equations using the measure units
\begin{eqnarray}
\sigma_{s}{=}\left( {\rho c_{\upsilon}\eta_{0}T_{c} \over
\tau _{T}}\right)^{1/2},\quad  \varepsilon_{s}{=}
{\frac{\sigma_{s}}{G_{0}}}{\equiv} \left( {\frac{\tau _{\varepsilon
}}{\tau _{T}}}\right) ^{1/2}\left( {\frac{\rho c_{\upsilon}T_{c}\tau
_{\varepsilon } }{\eta_{0}}}\right)^{1/2},\quad T_{c}
\label{1a} \end{eqnarray}
for variables $\sigma$, $\varepsilon $, $T$, respectively, where
$\rho$ is the mass density, $c_v$ is the specific heat capacity,
$T_{c}$ is the critical temperature,
$\eta_{0} \equiv \eta(T=2T_{c})$ is the typical value of shear viscosity
$\eta$, $\tau_{T}\equiv\rho l^2 c_{\upsilon}/\kappa$ is the time of heat
conductivity, $l$ is the scale of heat conductivity, $\kappa$ is the
heat conductivity constant, $\tau_{\varepsilon}$ is the relaxation time of
matter strain, $G_{0}\equiv \eta _{0}/\tau _{\varepsilon}$:
\begin{eqnarray}
&&\tau _{\sigma}\dot{\sigma}=-\sigma + g\varepsilon , \label{1} \\
&&\tau_{\varepsilon }\dot{\varepsilon}=-\varepsilon + (T-1)\sigma ,
\label{2} \\ &&\tau _{T}\dot{T}=(T_{e}-T) - \sigma \varepsilon +
\sigma ^{2}. \label{3} \end{eqnarray}
Here the stress relaxation time $\tau_{\sigma}$, the temperature $T_{e}$
of atomically flat mica friction surfaces, and the
constant $g=G/G_{0}$ are introduced, where $G$ is the lubricant shear modulus.
It can be seen \cite{liq,voigt}
that Eqs.~(\ref{1}) and (\ref{2}) are the Maxwell-type and the Kelvin-Voigt
equations for viscoelastic matter, correspondingly.
The latter takes into account the dependence of the
shear viscosity on the dimensionless temperature $\eta = \eta_{0}/(T-1)$.
Equation~(\ref{3}) represents the heat conductivity expression,
which describes the heat transfer from the friction surfaces to the
layer of lubricant, the effect of the dissipative heating
of a viscous liquid flowing under the action of the stress, and the reversible
mechanic-and-caloric effect in linear approximation.
These equations coincide with the synergetic Lorenz system formally
\cite{Haken,zhetph}, where the elastic shear stress
acts as the order parameter, the conjugate field
is reduced to the elastic shear strain, and the temperature is the control
parameter. As is known this system can be used for description of
the thermodynamic phase and the kinetic transitions.

In Ref.~\cite{liq} a melting of ultrathin lubricant film by friction
between atomically flat mica surfaces has been represented as a result of
action of spontaneously appearing elastic field of stress
shear component caused by the heating of friction surfaces above
the critical value $T_c=1+g^{-1}$. Thus, according to such approach
the studied solid-liquid transition of lubricant film
occurs due to both thermodynamic and shear melting.
The initial reason for this self-organization process
is the positive feedback of $T$ and $\sigma$ on
$\varepsilon$ [see Eq.~(\ref{2})] conditioned by the temperature
dependence of the shear viscosity leading to its divergence.
On the other hand, the negative feedback of $\sigma$ and $\varepsilon$
on $T$ in Eq.~(\ref{3}) plays an important role since it ensures the system
stability.

According to this approach the lubricant represents a strongly viscous liquid
that can behave itself similar to the solid --- has a high effective
viscosity and still exhibits a yield stress \cite{Yosh,voigt}.
Its liquidlike state
corresponds to the elastic shear stress which relax to the
zero value during larger times $\tau_{\sigma}$ than that of the
solidlike state. Moreover, at $\sigma = 0$ Eq.~(\ref{2}) reduces to the
Debye law describing the rapid relaxation of the elastic shear strain
during the microscopic time $\tau_{\varepsilon}\sim 10^{-12}$ s.
At that the heat conductivity equation (\ref{3}) takes the form of
simplest expression for temperature relaxation that does not contain the
terms representing the dissipative heating and the mechanic-and-caloric
effect of a viscous liquid. Also, we assumed that the film becomes more 
liquidlike and the friction decreases with the temperature growth 
due to decreasing activation energy barrier to molecular hops.

In accordance with Ref.~\cite{Aranson} in the absence of shear
deformations the temperature mean-square
displacement is defined by equality $\langle u^2\rangle=T/Ga$, where
$a$ is the lattice constant or the intermolecular distance.
The average shear displacement is found from the relationship
$\langle u^2\rangle =\sigma^2a^2/G^2$. The total mean-square displacement
represents the sum of these expressions provided that the thermal
fluctuations and the stress are independent.
Above implies that the transition of lubricant from solidlike to
fluidlike state is induced both by heating and under influence of stress
generated by solid surfaces at friction. This agrees
with examination of solid state instability within the framework
of shear and dynamic disorder-driven melting representation in absence
of thermal fluctuations.
Thus, the strain fluctuations, related to the stress ones,
and the thermal fluctuations will be considered independently.

In present paper the additive noises of the shear components of
the elastic stress and strain, and the temperature are taken into account
in a lubricant film. The solidlike lubricant is
assumed to be amorphous (disordered). Therefore I study the glass
transition represented in terms of a second-order phase transition.
It is shown that increase of
the strain noise intensity causes the lubricant film melting even at
low temperatures of friction surfaces,
at which the temperature noise plays a crucial role.
The phase diagrams are calculated defining the domains of
sliding, stick-slip, and dry friction in the planes
temperature noise intensity --- temperature of friction surfaces and
noise intensity of shear elastic strain  --- temperature noise intensity.

\section{Langevin and Fokker-Planck equations}\label{sec:level2}

Consider now the affect of additive noises of the elastic stress and
strain shear components $\sigma$, $\varepsilon$, and the temperature $T$.
With this aim, one should add to right-hand side of
Eqs.~(\ref{1}) --- (\ref{3}) the stochastic terms
$I_\sigma^{1/2}\xi$, $I_\varepsilon^{1/2}\xi$, $I_T^{1/2}\xi$
(here the noise intensities $I_{\sigma}$, $I_{\varepsilon}$, $I_{T}$ are
measured in units of $\sigma_{s}^2$, 
$\varepsilon_{s}^{2}\tau_{\varepsilon}^{-2}$, 
$\left( T_c \kappa/l\right)^2$, correspondingly, and
$\xi(t)$ is the $\delta$-correlated stochastic function) \cite{13}.
Then, within the adiabatic approximation
$\tau_{\sigma} \gg \tau_{\varepsilon},~ \tau_{T}$,
equations (\ref{2}) and (\ref{3}) are reduced to the
time dependencies
\begin{eqnarray} \varepsilon(t)&=&\bar \varepsilon + \tilde
\varepsilon\xi(t),\quad T(t)=\bar T + \widetilde T\xi(t); \label{X1}\\
\bar \varepsilon &{\equiv}&
\sigma \left(T_{e} {-} 1 {+} \sigma^2\right) d(\sigma), ~
\tilde{\varepsilon}{\equiv}\sqrt{I_\varepsilon {+} I_T \sigma^2}
~d(\sigma), \nonumber \\
\bar T &{\equiv}&\left(T_{e} {+} 2\sigma^2\right) d(\sigma), ~
\widetilde{T}\equiv\sqrt{I_T {+} I_\varepsilon\sigma^2}~d(\sigma), ~
d(\sigma)\equiv(1+\sigma^2)^{-1}. \label{X2}
\end{eqnarray}
Here, deterministic components are reduced to obtained in Ref.~\cite{liq},
whereas fluctuational ones follow from the known property of
variance additivity of independent Gaussian random quantities
\cite{13}. Thus, using the slaving principle inherent in
synergetics \cite{Haken,zhetph} transforms noises of both strain
$\varepsilon$ and temperature $T$, which are adiabatic initially, to
multiplicative form.  As a result, a combination of
Eqs.~(\ref{1}), (\ref{X1}), and (\ref{X2}) leads to the Langevin
equation
\begin{equation} \dot \sigma = f(\sigma) + \sqrt{I(\sigma)}~\xi(t),
\quad f\equiv-~{\partial V\over\partial \sigma}, \label{VI.5_17a}
\end{equation}
where the force $f$ is related to the synergetic potential \cite{liq}
\begin{equation}
V={\frac{1}{2}}(1-g)\sigma ^{2}+g\left(1-{\frac{T_{e}}{2}}\right)
\ln(1+\sigma^{2}) \label{X19}
\end{equation}
and an expression for the effective noise intensity
\begin{equation}
I(\sigma)\equiv I_\sigma + \left(I_\varepsilon +
I_T\sigma^2\right) g^2 d^2(\sigma)
\label{X3}
\end{equation}
is obtained in accordance with above mentioned property of noise
variance additivity. In order to avoid mistakes, one should notice
that a direct insertion of Eqs.~(\ref{X1}) and (\ref{X2}) into
(\ref{1}) results in the appearance of a stochastic addition
\begin{equation}
\left[I_\sigma^{1/2} + \left(I_\varepsilon^{1/2} +
I_T^{1/2}\sigma\right)gd(\sigma)\right]\xi(t)
\label{X4}
\end{equation}
whose squared amplitude is quite different from the
effective noise intensity (\ref{X3}).
Moreover, in contrast to the expressions (\ref{X2}), a direct use of the
adiabatic approximation in Eqs.~(\ref{2}) and (\ref{3}) reduces the fluctuational
additions in Eqs.~(\ref{X1}) to the forms:
$\tilde{\varepsilon}\equiv(I_\varepsilon^{1/2}+I_T^{1/2}\sigma)
d(\sigma)$, $\widetilde T\equiv(I_T^{1/2} - I_\varepsilon^{1/2}
\sigma)d(\sigma)$.
The latter is obviously erroneous since the
effective noise of the temperature $\widetilde T$ disappears entirely
for the stress $\sigma=\sqrt{I_T/I_\varepsilon}$.
The reason for such a contradiction is that the Langevin equation
does not permit the use of usual analysis methods (see \cite{13}).

To continue in the usual way, let us write the Fokker-Planck equation
related to the Langevin Eq.~(\ref{VI.5_17a}):
\begin{equation}
{\partial P(\sigma, t)\over\partial t}{=}
{\partial\over\partial \sigma}\left\{{-}f(\sigma) P(\sigma, t){+}{\partial
\over\partial \sigma}\left[I(\sigma)P(\sigma, t)\right]\right\}.
\label{X5}
\end{equation}
At steady state, that is the single considered case,
the probability distribution $P(\sigma, t)$ becomes
a time-independent function $P(\sigma)$.
Consequently, under the usual condition, that
the expression in braces of the right-hand side of Eq.~(\ref{X5}) is
equal to zero, this leads to a stationary distribution
\begin{equation}
P(\sigma)=Z^{-1}\exp\lbrace-U(\sigma)\rbrace,
\label{a}
\end{equation}
where $Z$ is a normalization constant. The effective
potential
\begin{equation}
U(\sigma)=\ln I(\sigma)-\int\limits^\sigma_0{f(\sigma')\over I(\sigma')}{\rm d}\sigma',\quad
f\equiv -~{\partial V\over\partial \sigma},
\label{VI.5_17}
\end{equation}
is determined by the synergetic potential $V$ [Eq.~(\ref{X19})]
and the noise intensity $I(\sigma)$ [Eq.~(\ref{X3})] \cite{14}.
Combining these expressions, one can find the explicit form of $U(\sigma)$,
which is too cumbersome to be reproduced here.
The equation defining the locations of the maximums of the
distribution function $P(\sigma)$
\begin{eqnarray}
(1-g)x^3&{+}&g(2{-}T_e)x^2 {-} 2g^2 I_T x {+}
4g^2(I_T {-} I_\varepsilon) {=} 0, \quad x \equiv 1 + \sigma^2,
\label{VI.5_19}
\end{eqnarray}
is much simpler.
According to Eq.~(\ref{VI.5_19}), maximums are insensitive to changes
in the intensity of the noise $I_\sigma$ of the stress $\sigma$,
but they are determined by the value $T_e$
of the friction surfaces temperature and the intensities
$I_\varepsilon$ and $I_T$ of the noises of the strain $\varepsilon$
and the lubricant film temperature $T$, which acquire the multiplicative character
in Eq.~(\ref{X3}). Hence, for simplicity $I_\sigma$ can be set equal to $0$
and Eqs.~(\ref{X19}), (\ref{X3}), and (\ref{VI.5_17}) give
the following expression for the effective potential:
\begin{eqnarray}
&&U(\sigma)= {1 \over 2g^2I_T} \left\{ i \left[ i(1-g) - g \left(2 - T_e\right)
\right] \ln | 1+\sigma /i | + (1-g){\sigma^2\over 2} \right. \nonumber \\
&&+ \left.
\left[ g \left(2 - T_e \right) - i(1-g) \right] \sigma \right\} +
\ln \left[ g^2 d^2(\sigma) \left( I_\varepsilon {+}
\sigma^2 I_T \right) \right],\quad
i {\equiv} { I_\varepsilon \over I_T} {-} 1 . \label{VI.5_27}
\end{eqnarray}

\section{Phase diagrams}\label{sec:level3}

According to Eq.~(\ref{VI.5_19}),
the effective potential (\ref{VI.5_27}) has a minimum at $\sigma=0$
if the temperature $T_e$ does not exceed the critical level
\begin{equation}
T^c=1+g^{-1}+2g \left( I_T - 2 I_\varepsilon \right),
\label{VI.5_29}
\end{equation}
whose value increases at increasing the characteristic value of shear
viscosity $\eta_0$ and the temperature noise intensity,
but decreases with growth of the shear modulus $G$ of lubricant and
the strain noise. Here, the lubricant film does not melt.
The solutions of Eq.~(\ref{VI.5_19})
\begin{eqnarray}
\sigma^2_\pm &{=}& {1\over 2}
\left[ { g(T_e{-}2) \over 1{-}g} {-} 3 {\pm} \sqrt{
\left({g(T_e{-}2) \over 1{-}g}{+}7\right) \left( {g(T_e{-}2)\over 1{-}g}{-}1
\right) } \right] \label{VI.5_34}
\end{eqnarray}
are obtained on the line defined by
expression (\ref{VI.5_29}) after elimination of the root
$\sigma^2=0$. At $T_e {<} T_c^0 {=} 2(1{+}2g^{-1})/3$ the roots $\sigma_\pm$
are complex, starting from $T_e = T_c^0$ they become zero,
and at $T_e > T_c^0$
one has real magnitudes $\sigma_+= -\sigma_- \ne 0$
that implies lubricant film melting. If equality (\ref{VI.5_29}) is
fulfilled, the root $\sigma=0$ corresponds to the minimum of the
effective potential (\ref{VI.5_27}) at $T_e {<} T_c^0$, whereas at
$T_e > T_c^0$ this root corresponds to the maximum, and the roots
$\sigma_\pm$ --- to symmetrical minimums.

Now, let us find another condition for the stability of the roots
$\sigma_\pm$ in the simple case $I_\varepsilon=0$.
Setting the discriminant of Eq.~(\ref{VI.5_19}) equal to
zero, one gets the equations
\begin{eqnarray}
I_T=0,\quad I_T^2-I_T
\left\{ {27\over 2g} \left[ {1{-}g \over g} {+} {2{-}T_e \over 3} \right]
- {(2{-}T_e)^2 \over 8(1-g)}\right\} - {(2{-}T_e)^3 \over 2g(1-g)} = 0,
\label{VI.5_37}
\end{eqnarray}
the second of which gives
\begin{eqnarray}
2I_T = {27\over 2g} \left[ {1{-}g \over g} {+} {2{-}T_e \over 3} \right]
&-& {(2{-}T_e)^2 \over 8(1-g)} \nonumber \\  &\pm&
\left\{ \left[ {27\over 2g} \left( {1{-}g \over g} {+} {2{-}T_e \over 3}
\right) {-} {(2{-}T_e)^2 \over 8(1{-}g)}\right]^2 {+} {2(2{-}T_e)^3
\over g(1-g)} \right\}^{1/2}.
\label{VI.5_38}
\end{eqnarray}
This equation defines a bell-shaped curve $T_e(I_T)$, which intersects
the horizontal axis at the point
\begin{equation}
I_T {=} {9(3{-}4g){+}8g^2 \over 2g^2(1{-}g)} {+}
\left\{ \left[ {9(3{-}4g)+8g^2 \over 2g^2(1{-}g)}\right]^2 {+}
{16 \over g(1{-}g)} \right\}^{1/2},
\label{VI.5_38a}
\end{equation}
and vertical axis at the point $T_e=2$. It has a maximum $T_e=2g^{-1}$ at
\begin{equation}
I_T = { 2(1-g) \over g^{2}}. \label{VI.5_39}
\end{equation}
It is easy to see that line (\ref{VI.5_29}) touches the curve
(\ref{VI.5_38}) at the tricritical point
\begin{equation}
T_e =T_c^0 = {2 \over 3} (1{+}2g^{-1}),\quad I_T = { 1-g \over 6g^{2}}.
\label{VI.5_36}
\end{equation}
Thus, this point addresses to the appearance of real roots
$\sigma_\pm\ne 0$ (\ref{VI.5_34}) of Eq.~(\ref{VI.5_19}) that means
lubricant film melting.

Let us now consider the more general case of two
multiplicative noises $I_\varepsilon$, $I_T\ne 0$.
The condition of extremum of the effective potential (\ref{VI.5_27}) splits
into two equations, one of which is simply  $\sigma=0$, and the other one
is given by Eq.~(\ref{VI.5_19}). As mentioned above, the
analysis of the latter indicates
that the line of existence of the zero solution is defined by
expression (\ref{VI.5_29}). The tricritical point $T$ has the coordinates
\begin{equation}
T_e {=} {2 \over 3} (1{+}2g^{-1} {-} 2gI_\varepsilon ),  ~
I_T {=} {1 \over 6g} (g^{-1} {-} 1 {+} 8gI_\varepsilon ) .
\label{VI.5_48}
\end{equation}
The phase diagrams for the fixed intensities $I_\varepsilon$ are shown in
Fig.~1. Physically, one should take into consideration that lines $1$ and
$2$ define the thresholds of stability loss of the system.
Above straight line $1$ the system manifests a stable
sliding friction (SF) inherent in the liquidlike phase of lubricant,
below curve $2$ the dry friction (DF) occurs that is characteristic for the
solidlike state of lubricant film. Between these lines the region of the
stick-slip friction (SS) mode is realized, i.e., mode that is characterized
by periodic transitions between two dynamic states during steady-state
sliding. It is relevant to an intermittent
regime of lubricant melting, where a mixture of both solidlike and liquidlike
phases exists. For $I_\varepsilon< (1+g^{-1})/4g$ the situation is
generally the same as in the simple case $I_\varepsilon=0$ (see Fig.~1a). At
$I_\varepsilon > (1+g^{-1})/4g$ the sliding friction is possible
even for small values of
temperature $T_e$ of friction surfaces and  noise
intensities $I_T$ of the lubricant film temperature (Fig.~1b). According to
(\ref{VI.5_48}), the tricritical point lies on the $I_T$-axis
at $I_\varepsilon = (1+2g^{-1})/2g$, and if the noise intensity
$I_\varepsilon$ is larger than the critical value $I_\varepsilon {=}
2g^{-2}$, the stable dry friction domain disappears (see Fig.~1c).
It is worth noting that this domain decreases with increase of the
shear modulus value $G$ and decrease of the characteristic value of
shear viscosity $\eta_0$.

\begin{figure}[!ht]
\begin{center}
\includegraphics[width=140mm]{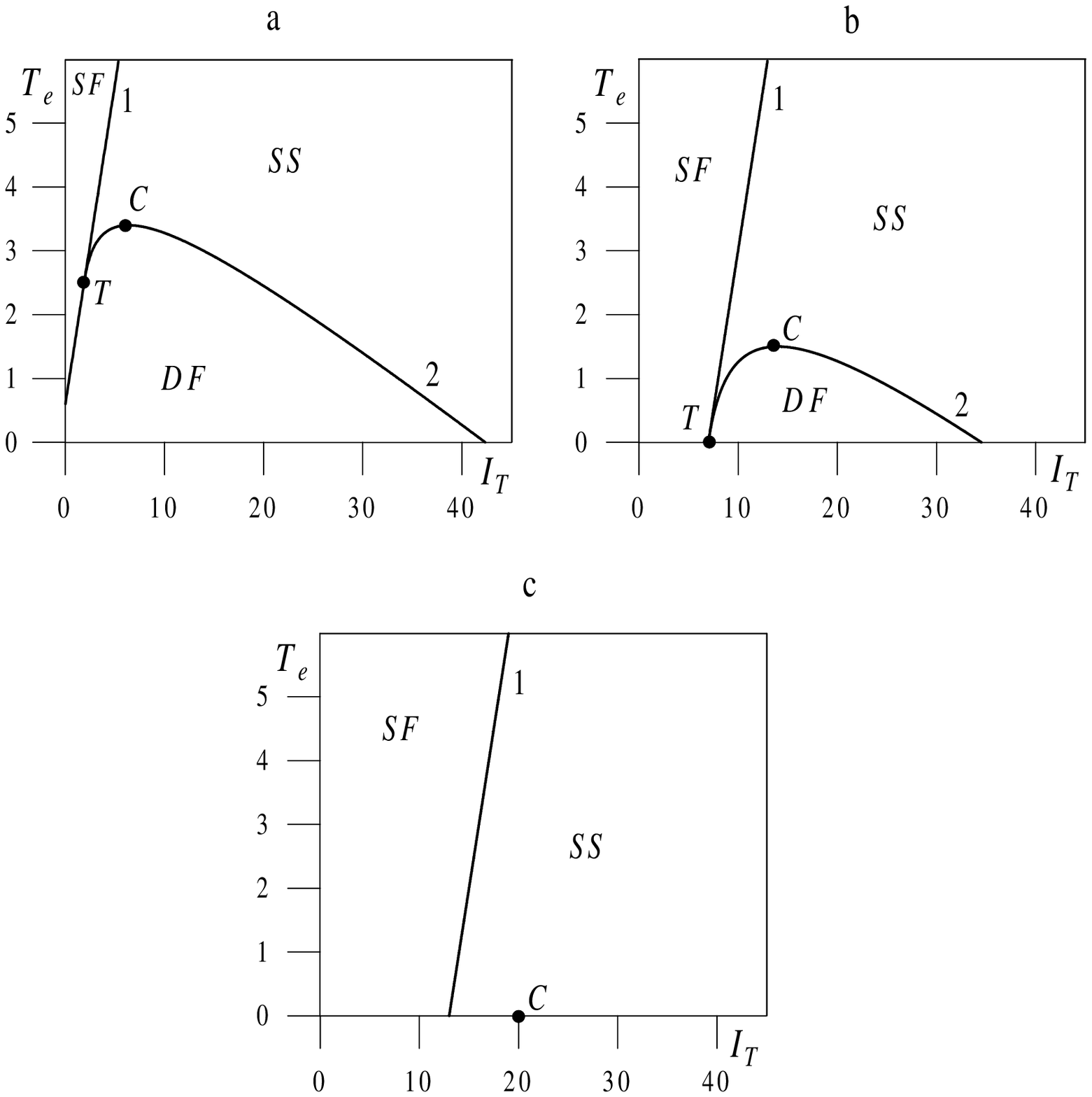}
\caption{Phase diagrams at $g{=}0.5$ and fixed values $I_\varepsilon$:
(a) $I_\varepsilon{=}1.2$; (b) $I_\varepsilon{=}5$; (c) $I_\varepsilon{=}8$.
Lines 1 and 2 define the boundary of stability domains of sliding (SF),
dry (DF), and stick-slip (SS) friction ($T$ is the tricritical point,
$C$ is the critical point).}
\end{center}
\end{figure}

The consideration of the additive noises of $\sigma$, $\varepsilon$, and $T$
shows that the stochasticity influence is non-essential
for the shear component of elastic stress tensor and it is crucial for
both the corresponding component of strain and the lubricant film temperature.
The boundary of the domain of sliding friction is fixed by the
equality for the noise intensities
\begin{equation}
I_T= 2 I_\varepsilon - (1+g)/2g^{2},
\label{bbb}
\end{equation}
following from Eq.~(\ref{VI.5_19}) at the conditions $x{=}1$ $(\sigma{=}0)$,
and $T_e{=}0$.
According to Eq.~(\ref{bbb}), in absence of the temperature noise
the lubricant melting occurs if the noise intensity of the shear strain
component exceeds the value
\begin{equation}
I_{\varepsilon 0} = (1+g)/4g^{2},
\label{b}
\end{equation}
corresponding to the point $O$ in Fig.~2.
The increase of both the shear strain and the temperature noises
causes the lubricant melting if their intensities are bounded
by condition (\ref{bbb}). The domain of the stick-slip friction appears
with further increase of these intensities above magnitudes
\begin{equation}
I_{\varepsilon 1} = {(2{+}g) \over 2g^2}, \quad
I_{T1} = {3{+}g \over 2g^2 }
\label{bb}
\end{equation}
at the tricritical point $T$ in Fig.~2. Such an intermittent behavior
is realized within the region located above straight line (\ref{bbb}) and
outside the curve that is determined by
\begin{eqnarray}
I_{\varepsilon } {=}
I_T\left[1{+}{ g \over 3(1{-}g)} \right] &{+}& {4g \over 27(1{-}g)^2}
\nonumber \\ &{-}&\left\{ { 2g^2 \over 27(1{-}g)}
\left[ { 4 \over 3(1{-}g)^2}\left({ 2 \over 9(1{-}g)}{+}I_T \right) {+}
{2I_T^2 \over 1{-}g} {+} I_T^3 \right] \right\}^{1/2}.
\label{ca}
\end{eqnarray}
If the noise intensity of the shear strain exceeds the value
$I_{\varepsilon 2}$ defined by (\ref{ca}) with the temperature noise
$I_{T2}{=} 2(3{-}g)/g^2$
(the point $C$ in Fig.~2), the dry friction region disappears at all.
The curve (\ref{ca}) intersects the vertical axis at the point
\begin{eqnarray}
I_{T3} {=} { 1\over 2} \left\{{27(1{-}g)\over 2g^2}{+}{9 \over g}{-}
{1 \over 2(1{-}g)} {+} \left[ \left( { 27(1{-}g) \over 2g^2} {+}
{9 \over g} {-} {1 \over 2(1{-}g)}\right)^{2} {+} {16 \over g(1{-}g)}
\right]^{1/2} \right\}
\label{ca1}
\end{eqnarray}
above that the dry friction does not take place.
The corresponding phase diagram depicted in Fig.~2
has a very non-trivial form (especially, within the domain
$I_{\varepsilon 1}{\leq} I_{\varepsilon}{\leq} I_{\varepsilon 2}$).

\begin{figure}[!ht]
\begin{center}
\includegraphics[width=140mm]{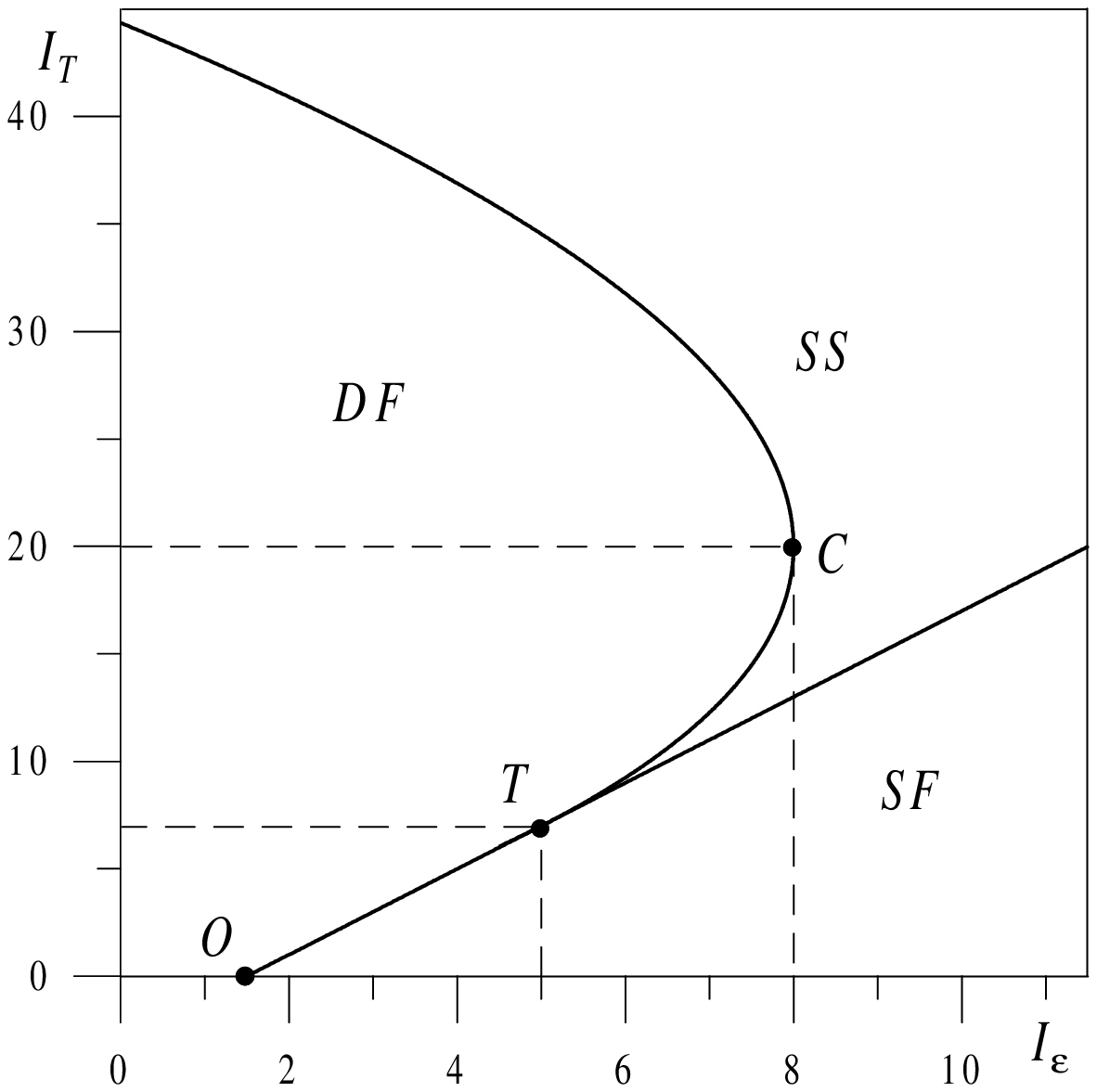}
\caption{\label{fig:epsart}Phase diagram for the system with
$T_e{=}0,~ g{=}0.5$, and $I_\varepsilon, ~I_T\ne 0$.}
\end{center}
\end{figure}

\section{Summary}\label{sec:level4}

The above consideration of the thermal and elastic
fields noise influence on the solid-liquid transition of ultrathin
lubricant film permits to define the domains of dry, sliding,
and stick-slip friction modes in the phase diagram. So that, an evidence
of the phase diagram complication is obtained due to studied
fluctuations. Depending on the initial conditions the growth of lubricant
film's temperature noise can decrease or increase friction, but the growth of
elastic shear strain noise increases the sliding friction region only.
It is shown that dry friction domain is bounded by relatively small values
of the confining walls temperature and the noise intensities of lubricant
strain and temperature (see Figs.~1~and~2). Thus, used here approach predicts
the possibility for controlling of frictional behavior.

Above the concept of dynamical shear melting of the ultrathin
lubricant film has been used \cite{5lit,Aranson}. 
In accordance with it the stick-slip friction can be described and
such melting is represented as a result of action of
elastic field of shear stress component caused by the heating of
friction surfaces above the critical value.
The essential limitation of this approach is the fact that stick-slip
motion I studied was independent of the way in which the system was driven,
i.e., elasticity and mass of the confining walls, although for such
friction mode the mentioned dependence is crucial.
It is worth noting that here the temperature of the confining walls
$T_e$ plays a role of the parameter of external influence.
Besides, the friction force decreases with
increasing velocity at the contact $V=l\partial\varepsilon/\partial t$
because the latter leads to the growth of the shear stress $\sigma$
according to the Maxwell-type stress - strain $\varepsilon$ relation:
$\partial\sigma /\partial t {= -}\sigma / \tau_\sigma {+} G\partial\varepsilon/
\partial t.$

\section*{Acknowledgment}

This work is dedicated to memory of my friend and colleague Dr. E.A.~Toropov.  I am grateful to Prof. A.I.~Olemskoi for helpful
suggestions, Dr.~I.~Krakovsky and Dr.~M.~Marvan for fruitful
discussions and hospitality during stay in Charles University,
Prague, and I.~Lyashenko for help with the numerical analysis of
Eq.~(\ref{VI.5_19}) and graphics. The work was partly
supported by the grant of Ministers Cabinet of Ukraine.

\end{document}